\documentclass[showkeys,showpacs,amsmath,amssymb,superscriptaddress,preprintnumbers,nofootinbib]{revtex4}
\usepackage{amssymb}
\usepackage{amsmath}
\usepackage{graphicx}
\usepackage{euscript}

\normalbaselines
\begin{document}

\title{Fingerprints of the Cosmological Constant: Folds in the Profiles of the Axionic Dark Matter Distribution in a Dyon Exterior}

\author{Alexander B. Balakin}
\email{Alexander.Balakin@kpfu.ru}
\author{Dmitry E. Groshev}
\email{groshevdmitri@mail.ru}
\affiliation{Department of General Relativity and
Gravitation, Institute of Physics,Kazan Federal University, Kremlevskaya street 18, 420008, Kazan,
Russia}

\newcommand{\nablab}{{\mathop {\rule{0pt}{0pt}{\nabla}}\limits^{\bot}}\rule{0pt}{0pt}}

\begin{abstract}
We consider the magnetic monopole in the axionic dark matter environment (axionic dyon) in the framework of the Reissner-Nordstr\"om-de Sitter model. Our aim is to study the distribution of the pseudoscalar (axion) and electric fields  near the so-called folds, which are characterized by the profiles with the central minimum, the barrier on the left, and the maximum on the right of this minimum. The electric field in the fold-like zones is shown to change the sign twice, i.e., the electric structure of the near zone of the axionic dyon contains the domain similar to a double electric layer. We have shown that the described fold-like structures in the profile of the gravitational potential, and~in the profiles of the electric and axion fields can exist, when the value of the dyon mass belongs to the interval enclosed between two critical masses, which depend on the cosmological constant.
\end{abstract}



\maketitle

\section{Introduction}

Global and local phenomena in our Universe are interrelated. When we speak about the cosmological constant $\Lambda$, first of all we think about the global structure of the Universe and about the rate of its expansion~\cite{L0}. On~the other hand, the~cosmological constant is associated with one of the models of the dark energy~\cite{L1,L2,L3}, and~the influence of the $\Lambda$ term on the structure of compact objects, black holes and wormholes can be interpreted as the dark energy fingerprints~\cite{BZworm}. Specific details of the causal structure of spherically symmetric objects, the~number and types of the horizons can also be associated with the cosmological constant. For~instance, in~the framework of non-minimal extensions of the Einstein--Maxwell and Einstein--Yang--Mills models with non-vanishing cosmological constant the solutions regular at the center appear, if~the parameters of the non-minimal coupling are linked by specific relationships with the cosmological constant~\cite{BLZ1,BLZ2}. In~other words, the~regularity at the center is connected with the appropriate asymptotic~behavior.

The dark matter, the~second key element of all modern cosmological scenaria~\cite{DM1,DM2,DM3}, can play the unique role in the problem of identification of the relativistic compact objects with strong electromagnetic fields. When one studies the properties of the gravitational field of the object, the~standard theoretical tool is based on the analysis of the dynamics of test particles; usually, one studies the trajectories of massive and massless particles reconstructing the fine details of the gravitational fields. When we deal with the axions, the~light pseudo-bosons~\cite{A1,A2,A3}, as~the possible representatives of the dark matter~\cite{ADM01,ADM02,ADM03,ADM1,ADM2,ADM3,ADM4,ADM5}, we can not monitor the motion of an individual particle, and~the analysis of the dark matter halos comes to the~fore.

The dark matter particles are not yet identified. There are few candidates, which can be classified as WIMPs and non-WIMPs (Weekly Interacting Massive Particles); the sets of candidates include light bosons and fermions; these candidates can compose the systems indicated as cold, warm and hot dark matter components. In~fact, one can assume that there are a few different fractions, which are united by a common term dark matter.
We analyze the axionic fraction of the dark matter, and~below we use the short terms {axionic dark matter} and {axions}. For~the description of the axionic dark matter, we use the master equations for the interacting pseudoscalar and electromagnetic field~\cite{WTN}; over the last thirty years, they have been known as the equations of axion electrodynamics~\cite{Sikivie,Wil}. Furthermore, we follow the idea that the axionic fraction of the dark matter behaves as a correlated system, in~particular, the~axions can form the Bose--Einstein condensate~\cite{BEC}.

 The dark matter forms specific cosmological units: halos, sub-halos, filaments, walls, the~structure of which admits the recognition  of the type of the corresponding central elements. We assume that compact relativistic objects with strong gravitational and electromagnetic fields distort the axionic fraction of the dark matter halos, which surround them (see, e.g.,~\cite{BG2,BG3} for details). For~instance, when the magnetic field of the star possesses the dipolar component, the~axionic halo is no longer spherically symmetric and is characterized by dipole, quadruple, etc. moments.  In~other words, the~modeling of the halo profiles for the magnetic stars can be useful in the procedure of identification of these objects, as~well as, in~the detailing of their~structures.

Our goal is to analyze specific details of the axionic dark matter profiles, which can be formed near the folds in the profile of the gravitational potential. The~axion distribution near the folds is non-monotonic, thus the electric field induced by the strong magnetic field in the axionic environment can signal about the appearance of the inverted layers analogous to the ones in the axionically active plasma~\cite{BG1}. We show that, when the spacetime is characterized by the non-vanishing cosmological constant, the~folds in the profiles of the gravitational potential appear, the~solutions to the equations for the pseudoscalar (axion) field inherit the fold-like behavior, and~the axionically induced electric field changes the sign in these domains in analogy with the phenomena of stratification in the axionically active plasma~\cite{BG1}.

The paper is organized as follows. In~Section~\ref{sec2}, we discuss mathematical details, describing the magneto-electro-statics  of the axionic dyons. In~Section~\ref{sec3}, we introduce and mathematically describe the idea of folds in the profile of the gravitational potential. In~Section~\ref{sec4}, we analyze the solutions to the key equations of the model and illustrate the appearance of the fold-like structures in the profiles of axion and electric fields. Section~\ref{sec5} contains~discussion.

\section{Description of the Axionic Dark Matter~Profiles\label{sec2}}

\subsection{The Total Action~Functional}

We consider the model, which can be described by the  total action functional
\begin{equation}
S_{(\rm total)} = S_{(\rm EH)} + S_{(\rm BMF)} + S_{(\rm AE)} \,.
\label{1}
\end{equation}
Here $S_{(\rm EH)}$ is the Einstein--Hilbert functional with the Ricci scalar $R$ and cosmological constant $\Lambda$
\begin{equation}
S_{(\rm EH)} = \int d^4x \sqrt{-g} \ \frac{(R+2\Lambda)}{2\kappa} \,,
\label{2}
\end{equation}
and $S_{(\rm BMF)}$ describes the contribution of a matter and/or fields, which form the background spacetime.
The action functional of the axiono-electromagnetic subsystem is represented in the form
\begin{equation}
S_{(\rm AE)} = \int d^4x \sqrt{-g} \left\{ \frac{1}{4}F^{mn}\left(F_{mn}+\phi F^*_{mn} \right) + \frac12 \Psi^2_0 \left[V - \nabla_m \phi \nabla^m \phi \right] \right\} \,,
\label{4}
\end{equation}
where $F_{mn}$ is the Maxwell tensor and  $F^*_{mn}$ is its dual tensor; $\phi$ denotes the dimensionless pseudoscalar (axion) field, $V$ is the potential of the pseudoscalar field, and~$\Psi_0 = \frac{1}{g_{A \gamma \gamma}}$ is the parameter reciprocal to the constant of the axion-photon coupling $g_{A \gamma \gamma}$.

\subsection{Background~State}

We follow the hierarchical approach, according to which the background gravitational field is considered to be fixed, and~the axionic dark matter is distributed in this given spacetime. Why~we do it? The relativistic objects of the neutron star type are compact, but~the mass density inside these objects is very high, about $\rho_{\rm n} \propto 10^{15} {\rm g}/{\rm cm}^3$.  The~average mass density of the dark matter is known to be estimated as $\rho_{\rm DM} \propto 10^{-24} {\rm g}/{\rm cm}^3$, but, in~contrast to the dense objects, the~dark matter is distributed quasi-uniformly in the whole Universe. Thus, the~gravitational field in the vicinity of the dense magnetic stars is predetermined by the baryonic matter and by the magnetic field with very high energy. From~the mathematical point of view, in~order to describe the background state we use only two elements of the total action functional (\ref{1}), namely, $S_{(\rm EH)} {+} S_{(\rm BMF)}$, and~consider the known solutions to the corresponding master equations. As~for the axionic and electric subsystems, we obtain the master equations and analyze their solutions with the assumption that the gravity field is already~found.

\subsection{Master Equations of the Axion~Electrodynamics}

In this work we assume that the potential of the axion field is of the form $V(\phi)= m^2_{\rm A} \phi^2$. More~sophisticated periodic potential is considered in the papers~\cite{BG3,BG1}. Mention should be made, that we use the system of units, in~which $c{=}1$, $\hbar{=}1$, $G{=}1$; in this case the dimensionality of the axion  mass $m_{\rm A}$ coincides with the one of the inverse length. In~the standard system of units we have to replace $m_{\rm A}$  with $\frac{m_{\rm A} c}{\hbar}$.
Variation of the action functional (\ref{4}), with respect to pseudoscalar field $\phi$ gives the known master equation
\begin{equation}
\nabla_m \nabla^m \phi + m^2_{\rm A} \phi = - \frac{1}{4\Psi^2_0} F^{*}_{mn}F^{mn} \,.
\label{5}
\end{equation}
Variation procedure associated with the electromagnetic potential $A_i$ gives the equation
\begin{equation}
\nabla_k \left[F^{ik} + \phi F^{*ik} \right] =0 \,.
\label{7}
\end{equation}
This equation, being supplemented by the equation
\begin{equation}
\nabla_k F^{*ik}  =0 \,,
\label{8}
\end{equation}
can, as~usual, be transformed into
\begin{equation}
\nabla_k F^{ik} = -   F^{*ik} \nabla_k \phi \,.
\label{9}
\end{equation}
The Equations~(\ref{9}), (\ref{8}) and (\ref{5}) are known as the
master equations of the axion electrodynamics~\cite{WTN,Sikivie}.

\subsection{Static Spacetime with Spherical~Symmetry}

We assume that the background spacetime is static and spherically symmetric, and~is described by the metric
\begin{equation}
ds^2 = N(r) dt^2 - \frac{dr^2}{N(r)} - \frac{1}{r^2}\left(d \theta^2 + \sin^2{\theta} d\varphi^2 \right)  \,.
\label{10}
\end{equation}
When the pseudoscalar and electromagnetic  fields inherit the spacetime symmetry, we obtain that $\phi$ is the function of the radial variable only, $\phi(r)$, and~the potential of the electromagnetic field can be presented in the form
\begin{equation}
A_i = \delta_i^0 A_0 + \delta_i^{\varphi} A_{{\varphi}}  \,.
\label{107}
\end{equation}
When one deals with the magnetic monopole of the Dirac type, the~azimuthal component of the potential is considered to be chosen in the form
\begin{equation}
A_{{\varphi}}= Q_{\rm m} (1{-}\cos{\theta}) \,,
\label{109}
\end{equation}
where $Q_{\rm m}$ is the magnetic charge (see, e.g.,~\cite{Wil,Lee}). The~Equations~(\ref{9}) and (\ref{8}) can be reduced to one equation only, which contains the electrostatic potential $A_0(r)$
\begin{equation}
\frac{d}{dr}\left(r^2 \frac{d A_0}{dr} + Q_{\rm m} \phi \right) = 0  \,.
\label{11}
\end{equation}
The Equation~(\ref{5}) takes now the form
\begin{equation}
\frac{1}{r^2}\frac{d}{dr}\left(r^2 N \frac{d \phi}{dr} \right) - m^2_{\rm A} \phi = -  \frac{Q_{\rm m}}{\Psi^2_0 r^2} \left(\frac{d A_0}{dr}\right) \,.
\label{12}
\end{equation}
Integration of the Equation~(\ref{11}) gives
\begin{equation}
\frac{d A_0}{dr}   = \frac{{\cal Q}(r)}{r^2} \,, \quad {\cal Q}(r) \equiv K - Q_{\rm m} \phi \,,
\label{13}
\end{equation}
where $K$ is the constant of integration. The~function ${\cal Q}(r)=K {-} Q_{\rm m} \phi(r)$
plays here the role of an effective electric charge, which is virtually distributed around the object at the presence of the axion field. This idea allows us to use the term {axionically induced} electric~field.

As the next step, we replace the term $\frac{d A_0}{dr}$ in the right-hand side of (\ref{12}) with (\ref{13}), and~obtain the master equation for the pseudoscalar (axion) field
\begin{equation}
\frac{1}{r^2}\frac{d}{dr}\left(r^2 N \frac{d \phi}{dr} \right)  = \left[m^2_{\rm A} + \frac{Q^2_{\rm m} }{\Psi^2_0 r^4} \right]\phi - \frac{Q_{\rm m} K }{\Psi^2_0 r^4}  \,.
\label{14}
\end{equation}
Below, we analyze the solutions to the Equation~(\ref{14}) for models, in~which the known metric function $N(r)$ contains the non-vanishing cosmological~constant.

\section{On the Features of the Exact Solution Describing the Dirac Magnetic Monopole in the Spacetime with Cosmological~Constant\label{sec3}}

\subsection{Geometrical Aspects and Definition of the~Fold}

The magnetic monopole forms the background spacetime with the well-known metric
\begin{equation}
N=1-\frac{2M}{r} + \frac{Q^2_{\rm m}}{r^2} - \frac{1}{3} \Lambda r^2
\,.
\label{rn11}
\end{equation}
Since we work with the units with $c=1$ and $G=1$, the~asymptotic mass $M$  and the magnetic charge of the monopole $Q_{\rm m}$ have the formal dimensionality of the length. The~metric (\ref{rn11}) covers the following exact~solutions.

\vspace{2mm}
\noindent
1. When $Q_{\rm m}=0$ and $M=0$, we obtain the de Sitter metric in the so-called ${\cal R}$-representation.
\begin{equation}
N(r)= 1 - \frac{\Lambda}{3} r^2 \,.
\label{17}
\end{equation}
It is well-known that using the coordinate transformations
\begin{equation}
t=\tau {-} \frac12 \sqrt{\frac{3}{\Lambda}} {\rm ln}\left(1 {-} \frac{\Lambda}{3} R^2 \cdot e^{2\sqrt{\frac{\Lambda}{3}}\tau} \right) \,, \quad
r= e^{\sqrt{\frac{\Lambda}{3}}\tau } \cdot R \,,
\label{175}
\end{equation}
one can obtain the de Sitter metric in the ${\cal T}$ form:
\begin{equation}
ds^2 = {d\tau}^2 - e^{2\sqrt{\frac{\Lambda}{3}}\tau} \left[{dR}^2 + R^2\left({d\theta}^2 + \sin^2{\theta} {d \varphi}^2  \right) \right]\,.
\label{176}
\end{equation}
When the cosmological constant is positive, $\Lambda>0$, the~coordinate transformation (\ref{175}) is defined for the domain, in~which the argument of the logarithm is positive, i.e.,~when $r<r_{\rm H} \equiv \sqrt{\frac{3}{\Lambda}}$, in~other words, $r_{\rm H}$ indicates the cosmological horizon. When $\Lambda<0$, we deal with the anti-de Sitter spacetime, which has no~horizons.

\vspace{2mm}

\noindent
2. When $Q_{\rm m} = 0$, we obtain the Schwarzschild-de Sitter metric
\begin{equation}
N(r)= 1 -\frac{2M}{r} - \frac{\Lambda}{3} r^2 \,.
\label{177}
\end{equation}
When $\Lambda>0$, depending on the value of the mass $M$, the~spacetime can have two, one or zero horizons.
When $\Lambda<0$, there is one horizon. When $\Lambda$ {=} 0, we deal with the Schwarzschild model, which is characterized by one event horizon at $r$ {=} 2$M$.

\vspace{2mm}
\noindent
3. General case: The Schwarzschild--Reissner--Nordstr\"om--de Sitter~solution.

Searching for the horizons from the equation $N(r)=0$, we happen to be faced with the algebraic equation of the fourth order, and~this spacetime is known to be equipped by three horizons as maximum.
But our goal is more detailed: we would like to find the sets of parameters $M$, $Q_{\rm m}$ and $\Lambda$, for~which the profile $N(r)$ contains {\it folds}. We define the fold as the domain, which is characterized by the following two features:

\begin{itemize}

\item
the profile $N(r)$ has the central minimum, the~barrier on the left of the minimum, and~the maximum on the right;

\noindent
\item
this domain is inside the cosmological horizon, but~it is not hidden inside the event horizon, i.e.,~$N(r)>0$ in this~domain.

\end{itemize}

We have to stress, that when $\Lambda$ {=} 0, the~metric function $N(r)$ can have the barrier on the left, the~ minimum, but~there is no maximum on the right-hand side, there is only the monotonic curve, which tends asymptotically to the horizontal line $N~{=}~1$. Furthermore, one can imagine the folds of the second kind, for~which the barrier is on the right-hand side of the minimum, and~the maximum on the left-hand side, respectively. In~this paper we consider the first variant of the fold~only.

\subsection{Horizons}
\vspace{-6pt}

\subsubsection{Auxiliary Function Indicating the Number of~Horizons}

The analysis of the causal structure and of the folds appearance is based on the following approach (this approach was successfully applied in~\cite{BLZ1,BLZ2} for the case with the equation of the sixth order). First, we consider that $\Lambda >0$, and~rewrite the equation $N(r)=0$ in the form
\begin{equation}
M = \frac12 f(r) \,, \quad f(r)\equiv r + \frac{Q^2_{\rm m}}{r}- \frac{\Lambda}{3} r^3 \,.
\label{R1}
\end{equation}
The auxiliary function $f(r)$ starts from infinity at $r = 0$ and tends to minus infinity at $r \to \infty$;
it can possess two or zero local extrema depending on the value of the dimensionless guiding parameter~$\Lambda Q^2_{\rm m}$.

\subsubsection{The Case $\Lambda Q^2_{\rm~m}<\frac14$}

When $\Lambda Q^2_{\rm m}<\frac14$, the~function $f(r)$ has minimum and maximum, respectively, on~the spheres, indicated as
\begin{equation}
r_{1} = \sqrt{\frac{1}{2\Lambda}\left(1 - \sqrt{1-4 \Lambda Q^2_{\rm m}} \right)} \,,
\quad r_{2} = \sqrt{\frac{1}{2\Lambda}\left(1 + \sqrt{1-4 \Lambda Q^2_{\rm m}} \right)} \,.
\label{R3}
\end{equation}
The sketch of this function is depicted on the Figure 1a.
The number of horizons is predetermined by the number of the intersection points of the horizontal mass line $y~{=}~M$ and of the graph of the function $y~{=}~\frac12 f(r)$.

\begin{figure}[h]
\centering
\begin{minipage}[h]{0.32\linewidth}
	\center{\includegraphics[width=1\linewidth]{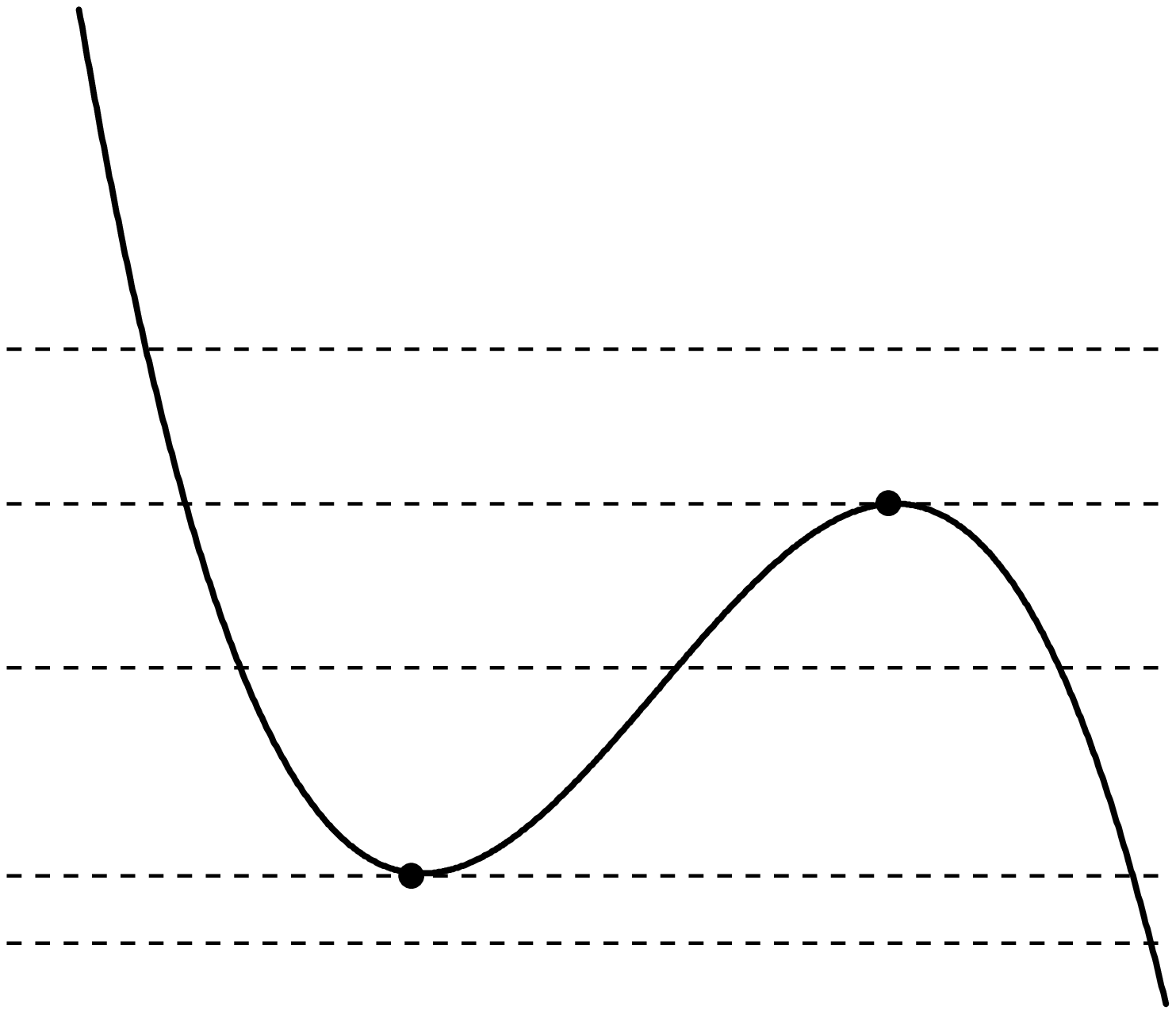}} (\textbf{a})
\end{minipage}
\begin{minipage}[h]{0.32\linewidth}
	\center{\includegraphics[width=1\linewidth]{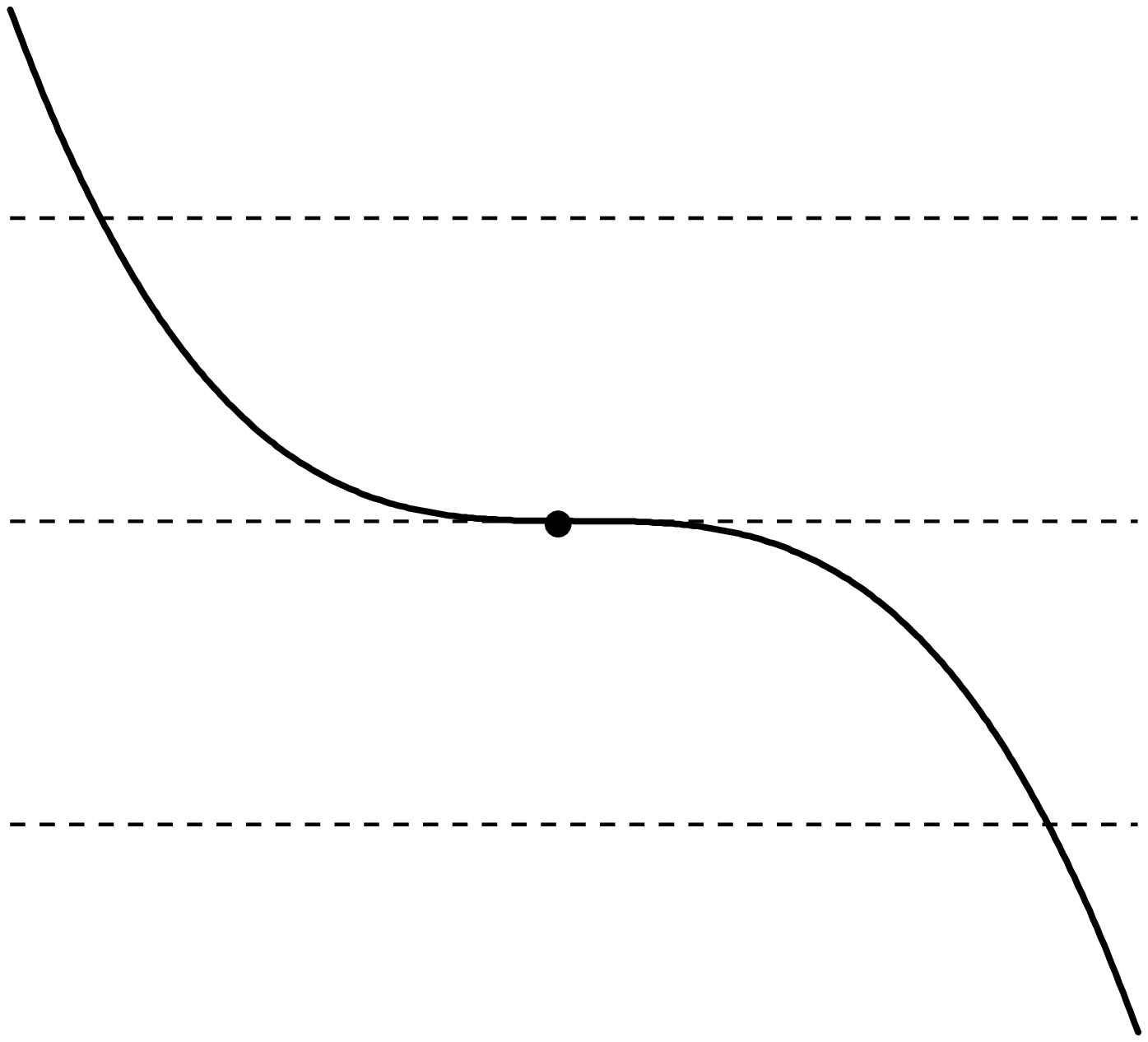}  (\textbf{b})}
\end{minipage}
\begin{minipage}[h]{0.32\linewidth}
	\center{\includegraphics[width=1\linewidth]{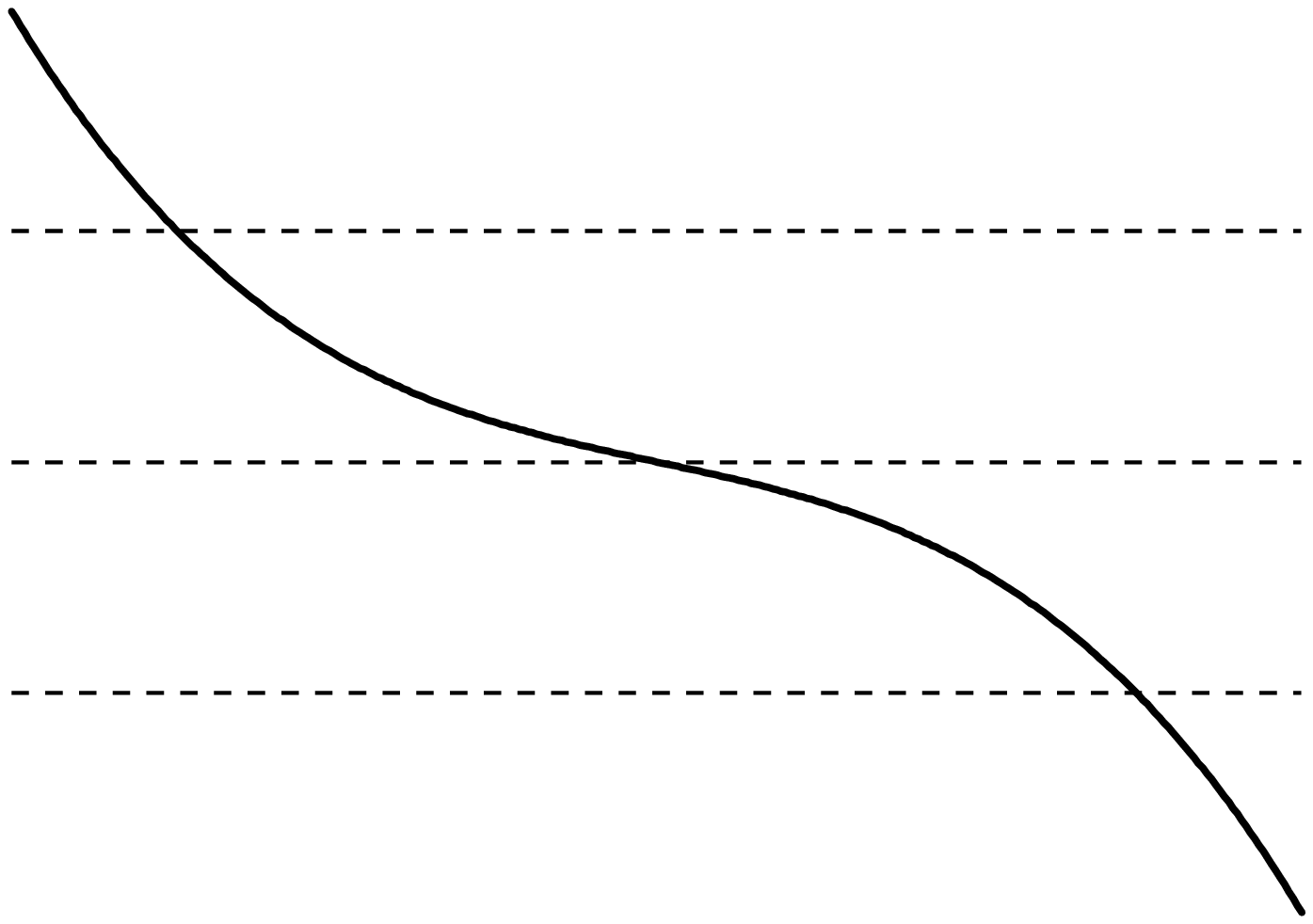}  (\textbf{c})}
\end{minipage}
\caption{Typical sketches of the auxiliary function (\ref{R1}), which illustrate the number of horizons depending on the values of the guiding parameter $\Lambda Q^2_{\rm m}$ and of the asymptotic mass $M$. Panel $(\textbf{a})$ illustrates the case $\Lambda Q^2_{\rm m}< \frac14$; panel $ (\textbf{b})$ relates to the case $\Lambda Q^2_{\rm m}=\frac14$, and~panel $ (\textbf{c})$ corresponds to $\Lambda Q^2_{\rm m}>\frac14$.}\label{fig1}
\end{figure}
According to the sketch $(a)$, the~values $r_1$ and $r_2$  define two critical values of the mass:
\begin{equation}
M_{1,2} = \frac12 f(r_{1,2}) =  \frac{\left(1\mp \sqrt{1-4 \Lambda Q^2_{\rm m}}+ 4\Lambda Q^2_{\rm m}\right)}{\sqrt{18\Lambda \left(1 \mp \sqrt{1-4 \Lambda Q^2_{\rm m}} \right)}}
\,.
\label{R4}
\end{equation}
Let us analyze the ratio ${\cal H} \equiv \frac{M_1}{|Q_{\rm m}|}$ as a function of the dimensionless parameter $z=\sqrt{4 \Lambda Q^2_{\rm m}}$:
\begin{equation}
{\cal H}(z) = \frac{\sqrt2}{3z^2} \left(1-\sqrt{1-z^2} + z^2 \right) \sqrt{1+\sqrt{1-z^2}} \,, \quad
{\cal H}^{\prime}(z) = - \frac{\sqrt2 z}{\sqrt{1-z^2} \sqrt{1+\sqrt{1-z^2}}}
\,.
\label{R48}
\end{equation}
Clearly, we deal with  defined on the interval $0\leq z <1$ monotonic function, which takes it maximal value ${\cal H}(0)=1$ at $z=0$, and~tends to
${\cal H}(1)=\frac{2\sqrt2}{3}<1 $ at $z=1$. In~other words, when $0<\Lambda Q^2_{\rm m}<\frac14$, the~ratio $\frac{M_1}{|Q_{\rm m}|}$ does not exceed one, and~the critical mass $M_1$ belongs to the interval $\frac{2\sqrt2}{3}|Q_{\rm m}|<M_1<|Q_{\rm m}|$. In~the standard units we have to replace $M_1 \to \frac{GM}{c^2}$ and
$Q_{\rm m} \to \frac{\sqrt{G}Q_{\rm m}}{\sqrt{4\pi \varepsilon_0} c^2}$, where $\varepsilon_0$ is the vacuum permittivity; thus, the~condition $M_1<|Q_{\rm m}|$ reads $M_1<\frac{|Q_{\rm m}|}{\sqrt{4 \pi \varepsilon_0 G}}$.

For different values of the asymptotic mass of the object, $M$, we obtain the following~results.

\begin{itemize}
\item
When $M<M_1$, the~mass line crosses the indicated graph in one point, i.e.,~there is only one (cosmological) horizon.

\item
When $M=M_1$, the~mass line is the tangent one with respect to the minimum of this graph, thus, there are two horizons: the double event horizon and the simple cosmological~one.

\item
When $M_1<M<M_2$, there are three intersection points, thus, there are three horizons: the inner and outer event horizons and the cosmological~one.

\item
When $M=M_2$ there are two horizons: the simple event horizon and the double cosmological~one.

\item
When $M>M_2$, there is only one horizon; but in contrast to the case $M<M_1$, it is the specific case, when all the apparent Universe is inside the event horizon (see, e.g.,~\cite{Odin1,Odin2} for some analogy).
\end{itemize}

\subsubsection{The Case $\Lambda Q^2_{\rm~m} = \frac14$ }

For this case the values $r_1$ and $r_2$ coincide, $r_{1} = r_2 = \sqrt{\frac{1}{2\Lambda}}$; the critical masses also coincide, $M_1~{=}~M_2 = \frac13 \sqrt{\frac{2}{\Lambda}}$. Now, according to the panel $ (b)$ of Figure~\ref{fig1}, there are simple cosmological horizons for every $M \neq \frac13 \sqrt{\frac{2}{\Lambda}}$, and~the event horizons are absent. In~the case $M=\frac13 \sqrt{\frac{2}{\Lambda}}$ the graph of the function $f(r)$ has the cubic  inflexion point, when the minimum and maximum coincide; now one has triple horizon, i.e.,~the inner, outer horizons coincide with the cosmological~one.

\subsubsection{The Case $\Lambda Q^2_{\rm~m} > \frac14$ }

For this case the values $r_1$ and $r_2$ are complex, thus, the~extrema of the auxiliary function $f(r)$ are absent, and~according to the panel $ (c)$ of Figure~1. for every mass $M$ there is only one intersection point corresponding to the cosmological~horizon.

\subsubsection{Short resume}

If we search for the models with the cosmological horizon, but~without the event horizons, we can choose one of the following conditions:

\begin{itemize}
\item[1.] $\Lambda Q^2_{\rm m} < \frac14$, $M<M_1$;

\item[2.] $\Lambda Q^2_{\rm m} = \frac14$, $M<\frac13 \sqrt{\frac{2}{\Lambda}}$;

\item[3.] $\Lambda Q^2_{\rm m} > \frac14$.

\end{itemize}

We consider below the first~case.

\subsection{Folds}

The next point of our analysis is the study of the folds. We consider now the derivative of the metric function
\begin{equation}
N^{\prime}(r)= \frac{2M}{r^2} - \frac{2Q^2_{\rm m}}{r^3}- \frac{2\Lambda}{3} r \,,
\label{R17}
\end{equation}
and  rewrite the equation $N^{\prime}(r)=0$ as follows:
\begin{equation}
M = \tilde{f}(r) \,, \quad \tilde{f}(r) \equiv \frac{Q^2_{\rm m}}{r} + \frac{\Lambda}{3} r^3
\,.
\label{R40}
\end{equation}
When we consider the extrema of the auxiliary function $\tilde{f}(r)$, we obtain that there exists the minimum at $r{=}r_{*}{=} \left(\frac{Q^2_{\rm m}}{\Lambda}\right)^{\frac14}$, which corresponds to the critical value of the mass
\begin{equation}
M_{\rm c} = \frac43 |Q_{\rm m}| (\Lambda Q^2_{\rm m})^{\frac14}\,.
\label{R419}
\end{equation}
When $4\Lambda Q^2 \leq 1$, it is simply to show that $M_1 \geq M_{\rm c}$. Indeed, the~ratio $\frac{M_1}{M_{\rm c}}$ can be rewritten as follows:
\begin{equation}
\frac{M_1}{M_{\rm c}} = \mu(z) = \frac{1+z^2 - \sqrt{1-z^2}}{2 z^{\frac32} \sqrt{1-\sqrt{1-z^2}}}\,,
\label{R4190}
\end{equation}
where $z = \sqrt{4\Lambda Q^2_{\rm m}}$. This ratio is equal to one, when $z=1$, and~tends monotonically to infinity at $z \to 0$, in~other words, the~graph of the function $\mu(z)$ locates above the horizontal line $y=1$, when $0<z<1$.

Taking into account these features, we can state the~following.

\begin{itemize}
\item
When $M>M_{\rm c}$, the~horizontal mass line  $y=M$ crosses the graph of the function $\tilde{f}(r)$
twice; this means that there are two extrema: the minimum and~maximum.

\item
When $M=M_{\rm c}$, two extrema coincide forming the cubic inflexion~point.

\item
When $M<M_{\rm c}$, the~profile $N(r)$ is monotonic.
\end{itemize}

We are interested to study the case $M>M_{\rm c}$, since only this case relates to the presence of the fold, which we search for.  At~the same time we hope to find the solution without event horizons (only the cosmological horizon can exist). It is possible when $M_{1}>M>M_{\rm c}$.

\subsection{Final Remarks about the Features of the Spacetime~Geometry}

\subsubsection{The Choice of the Appropriate Scale for the Radial~Variable }

\begin{figure}[h]
\centering
	\includegraphics[width=120mm,height=80mm]{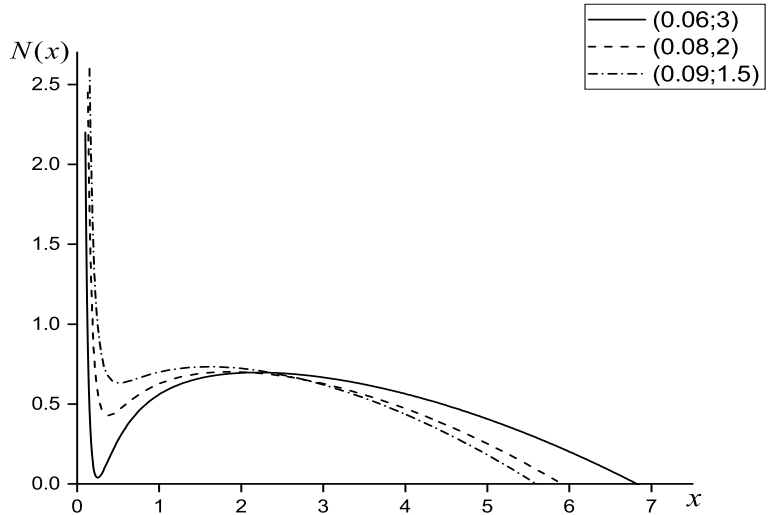}
	\caption{Folds in the profiles of the metric function $N(x)$ (\ref{R32}). There exists the infinite barrier on the left-hand side, the~central minimum, and~the maximum on the right-hand side. The~fold is situated in the domain with positive $N(x)$ and is not harbored by the event horizon. The~dimensionless guiding parameters $\sqrt{\Lambda Q^2_{\rm m}}$ and $\frac{M}{M_{\rm c}}$ are presented near the graphs in the~box.}\label{fig2}
\end{figure}

In order to clarify the possibility of the folds existence, we rewrite the basic equations using the replacement
\begin{equation}
r = \frac{x|Q_{\rm m}|}{(\Lambda Q^2_{\rm m})^{\frac14}} \,.
\label{R31}
\end{equation}
Keeping in mind that that there are two specific radii: first, the~quantity $r_{\Lambda}=\sqrt{\frac{1}{\Lambda}}$, which relates to the cosmological scale, second, $r_{\rm Q} = |Q_{\rm m}|$, which is the Reissner-Nordstr\"om radius, we see that (\ref{R31}) can be rewritten using the geometric mean value $r=x \sqrt{r_{\Lambda} r_{\rm Q}}$.
In terms of the new dimensionless variable $x$ we obtain
\begin{equation}
N(x)= 1+\sqrt{\Lambda Q^2_{\rm m}}\left[\frac{1}{x^2}- \frac{x^2}{3} - \frac{8}{3x} \left(\frac{M}{M_{\rm c}} \right) \right] \,.
\label{R32}
\end{equation}
This representation of the Reissner-Nordstr\"om-de Sitter metric shows explicitly that there are, in~fact, two dimensionless guiding parameters of the model: the parameter $\sqrt{\Lambda Q^2_{\rm m}}= \frac{r_{\rm Q}}{r_{\Lambda}}$ and the reduced mass $\frac{M}{M_{\rm c}}$. In~Figure~\ref{fig2}, we present three illustrations of the folds in the profile of the metric function $N(x)$, which has no event horizons, but~possesses the cosmological~horizon.

\subsubsection{On the Problem of Naked Singularity and Cosmic Censorship~Conjecture}

The point $r=0$ is singular for the metric with the coefficient (\ref{R32}). Since we consider the spacetime without event horizons, this point should be indicated as the naked singularity, which is typical for the Reissner--Nordstr\"om metric with dominating charge $Q_{\rm m}^2>M^2$ (the so-called super-extremal case). The~discussions about physical sense of such solutions were activated by Roger Penrose, who has formulated the cosmic censorship conjecture~\cite{C1}. Active debates concerning the physical status of the naked singularity continue until now, and~there are three groups of disputants. The~representatives of the first group insist that the solutions describing the naked singularity are non-physical. Scientists from the second group admit that such naked singularities exist in the nature, that it is still open problem, and~they propose the tests to verify this hypothesis (see, e.g.,~\cite{NS1,NS2,NS3,NS4} and references therein). We belong to the third group, and~we believe that the paradigm of the non-minimal coupling of the photons with the spacetime curvature removes this problem at all. In~the paper~\cite{BDZ} we have shown, that in the model of non-minimal Dirac monopole an additional horizon, formed due to the coupling of the photons to the curvature, appears, so that the point $r=0$ inevitably becomes hidden inside this non-minimal horizon; in other words, the~non-minimal interaction plays the role of the cosmic censor. Then in the work~\cite{BLZ1} we have found the exact solution to the non-minimally extended Einstein equations, which is regular at the center. For~that solution the metric coefficient $N(r)$ has the form
\begin{equation}
N(r)= 1+ \left(\frac{r^4}{r^4+ 2q Q^2_{\rm m}}\right)\left(-\frac{2M}{r} + \frac{Q^2_{\rm m}}{r^2} - \frac{1}{3} \Lambda r^2 \right) \,,
\label{Nregular}
\end{equation}
where $q$ is the non-minimal coupling parameter. The~solution (\ref{Nregular}) with $\Lambda=0$ was obtained  in~\cite{BZ2007}, and~then we studied it in~\cite{BZreg1,BZreg2}. Clearly, for~the solution (\ref{Nregular}) we obtain $N(0)=1$, i.e.,~the solution is regular at the center, and~the question about the naked singularity disappears. To~be more precise, we can attract attention of the Reader to the curve III in Fig.2 presented in the paper~\cite{BLZ1}: one can see the analog of the fold, but~the function $N(r)$ is regular at the center.
Why in the presented paper we considered the non-regular metric, if~we have the example of the regular one? The explanation is very simple: the non-minimal scale associated with the coupling parameter $q$ is estimated to be extremely small (of the order of the Compton radius of electron). This means that the folds, which we search for in the presented paper, are arranged rather far from the non-minimal zone, and~the metric (\ref{rn11}) gives the appropriate approximation for (\ref{Nregular}) in the fold~zone.

\section{Analysis of Solutions to the Key Equation of the Axion~Field\label{sec4}}

\subsection{The Profile of the Axion Field~Distribution}

In terms of the variable $x$ the key equation for the axion field (\ref{14}) takes the form
\begin{equation}
\phi^{\prime \prime}(x) + \phi^{\prime}(x)\frac{\left[x^2 N(x)\right]^{\prime}}{x^2 N(x)}  =
\frac{\phi |Q_{\rm m}|}{\sqrt{\Lambda} \Psi^2_0 x^4 N(x)} \left(m^2_{A} \Psi^2_0 x^4 {+} \Lambda \right) {-} \frac{K \sqrt{\Lambda} \ {\rm sgn}Q_{\rm m}}{\Psi^2_0 x^4 N(x)} \,.
\label{R317}
\end{equation}
Keeping in mind that the function $x^2 N(x)$ is the polynomial of the fourth order, and~that it has no zeros in the domain inside the cosmological horizon, $x<x_*$, we can see that this equation has only two singular points, $x=0$ and $x=x_*$, which are situated on the edges of the admissible interval $0<x<x_*$. Equation~(\ref{R317}) belongs to the class of the Fuchs equations~\cite{Fuchs}.

\begin{figure}[h]
\centering
	\includegraphics[width=120mm,height=80mm]{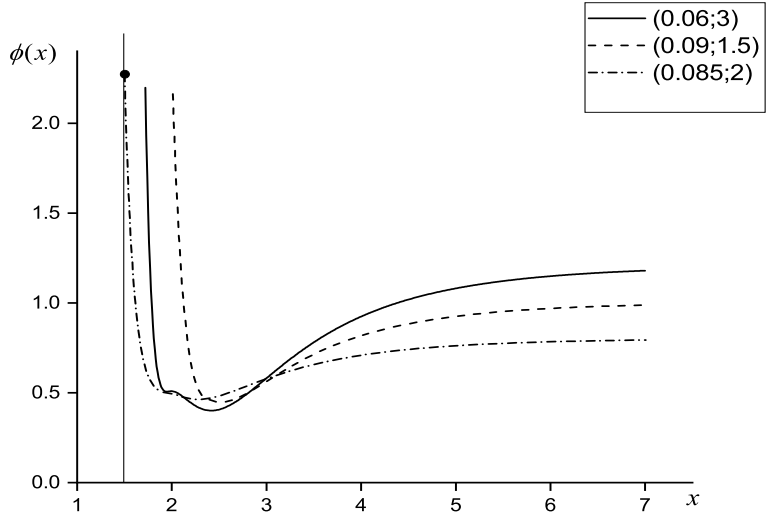}
\caption{Axion field profiles $\phi(x)$, as~the solutions to the master Equation~(\ref{R317}). The~guiding parameters of the model: $\sqrt{\Lambda Q^2_{\rm m}}$, and~$\frac{M}{M_c}$ are fixed in the box near the graphs; for the simplicity of illustration we put $\Psi_{0}=1$ and $m_{A}^{2}=0.1$ in the chosen system of units. The~vertical line symbolizes the delimiter associated with the boundary of the solid body of the object, and~its intersection with the graph $\phi(x)$ defines the boundary value $\phi(x_0)$. In~the far zone, the graph of the function $\phi(x)$ tends to the horizontal asymptotic line, which corresponds to $\phi_{\infty}$ given by (\ref{R717}). The~profiles of the axion field distribution inherit the fold-like structure of the profiles of the metric function $N(x)$. }\label{fig3}
\end{figure}

When $\Lambda>0$, we can not prolong the values of the radial variable to the infinity; we have to stop the analysis on the cosmological horizon $x=x_*$. In~other words, when we speak about the far zone, we mean the requirement $x \to x_*$, and~$N(x_*)=0$. One can see, that in this limit the axion field $\phi$ tends to the value $\phi_{\infty}$ given by
\begin{equation}
\phi_{\infty} = \frac{K \Lambda}{Q_{\rm m} \left(m^2_{A} \Psi^2_0 x^4_* + \Lambda \right)} \,.
\label{R717}
\end{equation}
The results for the near zone can be obtained numerically. Numerical simulation includes the variation of the following set of guiding parameters: first, the~parameter $\sqrt{\Lambda Q^2_{\rm m}}$ (it already appeared in the analysis of the function $N(x)$); second, the~parameter $m^2_{A}Q^2_{\rm m}$; third, the~parameter $\frac{K}{Q_{\rm m}}$, fourth, the~coupling constant $\Psi_0$. The~analysis has shown that the graphs $\phi(x)$ inherit the fold-like structure of the gravitational potential; for illustration, we presented three graphs in Figure~\ref{fig3}.

\subsection{The Profile of the Energy-Density of the Axionic Dark~Matter}

\begin{figure}[h]
\centering
	\includegraphics[width=120mm,height=80mm]{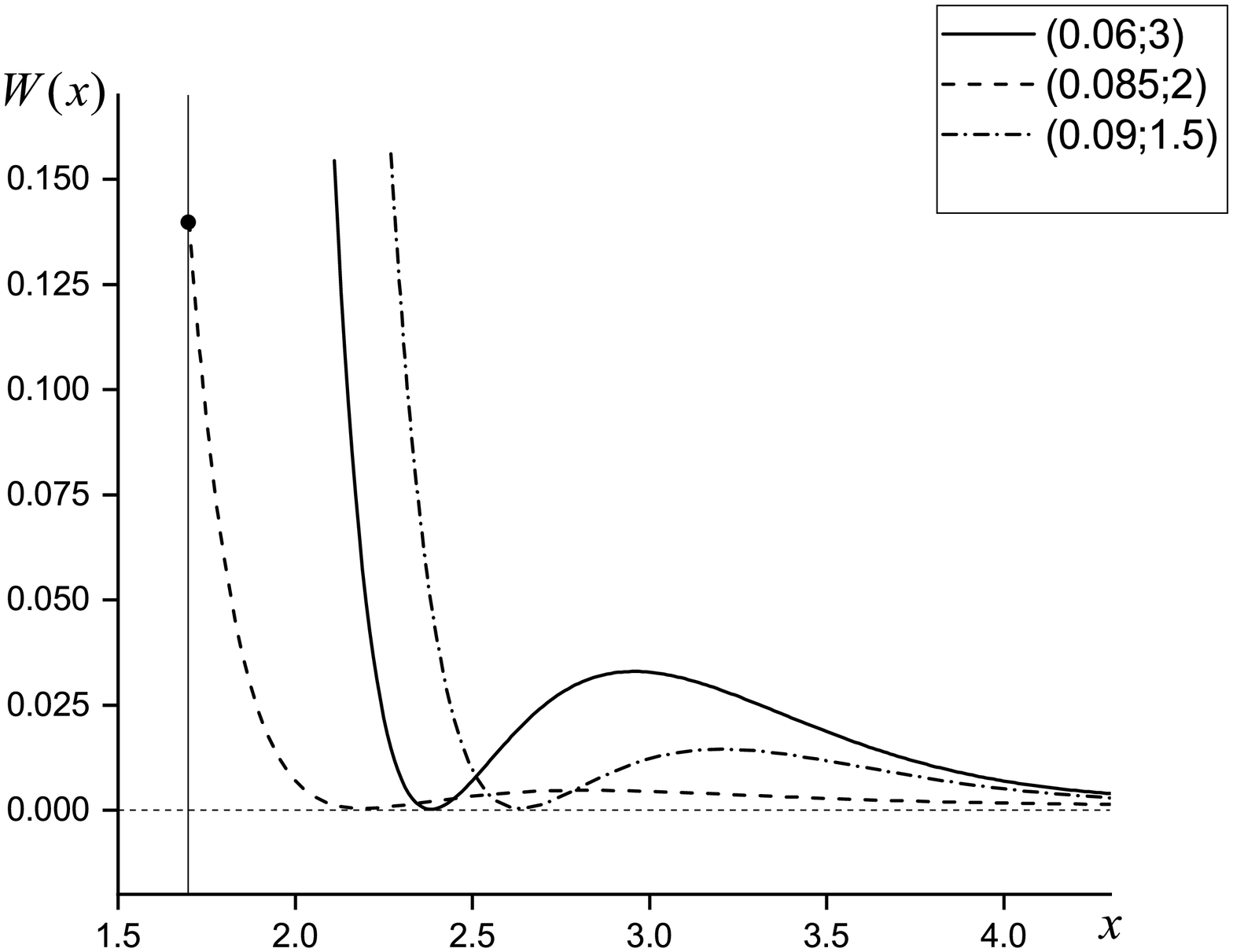}\vspace{-12pt}
\caption{Typical profiles of the energy density scalar of the axion field (\ref{W2}). The~basic profile has the typical fold-like structure: the minimum, the~barrier on the left of the minimum, the~maximum on its right-hand side. In~the far zone the axion energy density tends to a constant, and~the graphs have the horizontal asymptotes. The~vertical line relates to the object boundary and defines the corresponding boundary value $W(x_0)$.}\label{fig4}
\end{figure}

The scalar of the axion field energy density is standardly defined as
$W{=} U^i U^k T^{(\rm axion)}_{ik}$, where $U^i$ is the global four-velocity vector, coinciding with the normalized time-like Killing vector $\xi^i=\delta^i_0$. Normalization of this Killing four-vector gives $U^i= \frac{\xi^i}{\sqrt{\xi^s \xi_s}} = \frac{1}{\sqrt{N}} \delta^i_0$. The~quantity $T^{(\rm axion)}_{ik}$ is the stress-energy tensor of the pseudoscalar axion field. The~energy density scalar can be written as
\begin{equation}
W =  U^i U^k \Psi^2_0 \left\{\nabla_i \phi \nabla_k \phi {+} \frac12 g_{ik} \left[m^2_{A} \phi^2 {-} \nabla_n \phi \nabla^n \phi   \right]\right\} .
\label{W1}
\end{equation}
In the static spherically symmetric case we obtain
\begin{equation}
W(r) = \frac12 \Psi^2_0 \left[N(r) {\phi^{\prime}}^2(r) + m^2_{A} \phi^2\right]\,.
\label{W2}
\end{equation}
Using the profiles of the functions $N(x)$ and $\phi(x)$ we can illustrate the typical behavior of the profile of the function $W(x)$ (see Figure~\ref{fig4}). These profiles happen to be more sophisticated than the fold-like profile of the function $N(x)$, since the extrema  of the functions $N(x)$ and $\phi$ do not~coincide.

\subsection{Profiles of the Axionically Induced Electric~Field}

\begin{figure}[h]
\centering
	\includegraphics[width=100mm,height=80mm]{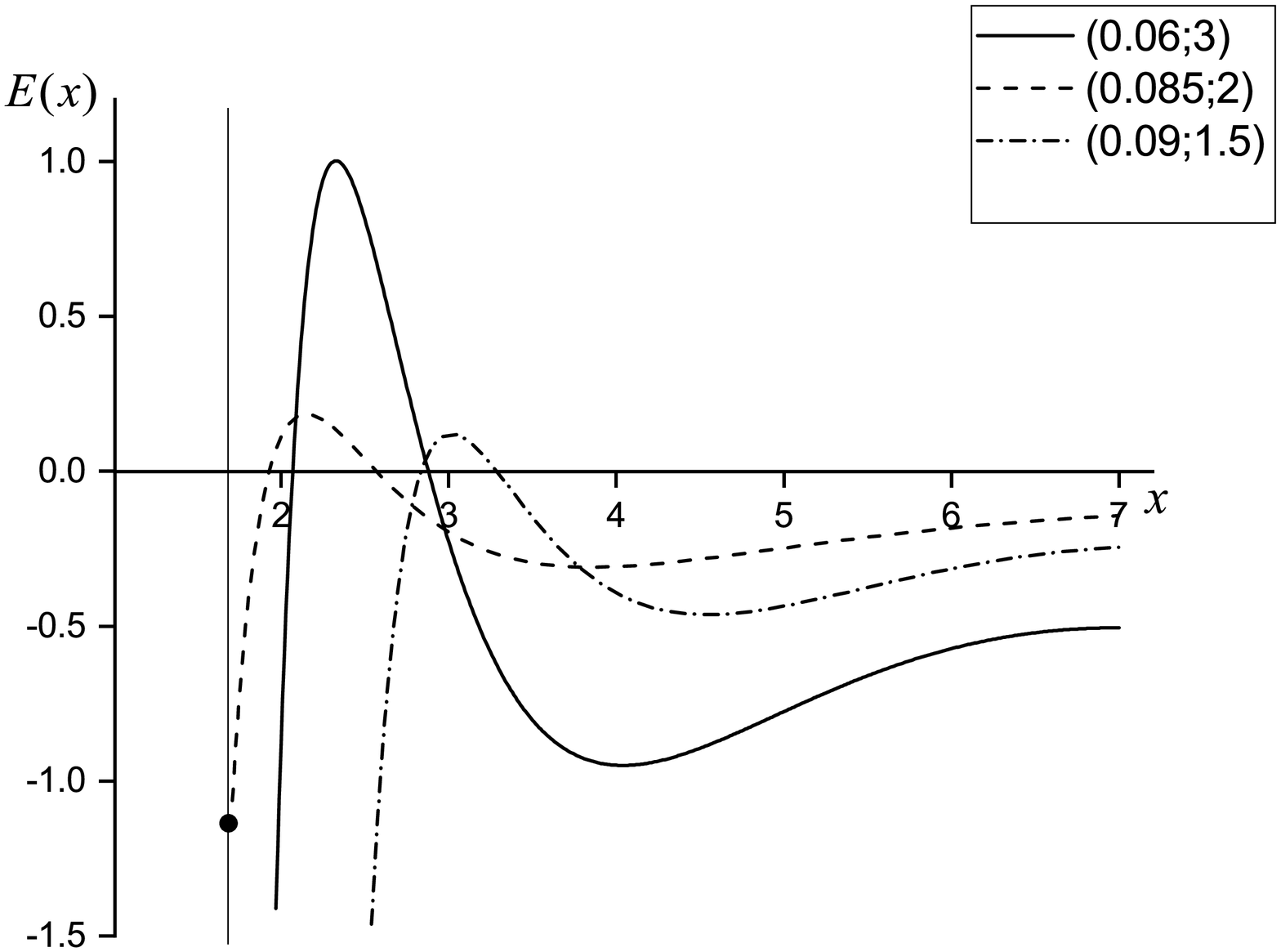}
\caption{Typical profiles of the axionically induced electric field. The~profiles have inverted fold-like structure. The~electric field changes sign twice; its profile tends to the Coulombian curve in the far zone near the cosmological horizon. The~vertical line relates to the boundary of the solid body of the object; the dot relates to the boundary value of the electric~field.}\label{fig5}
\end{figure}

When the profile of the axion field $\phi(x)$ is found, we can reconstruct the profile of the axionically induced electric field using the formula
\begin{equation}
E(x)= \frac{\sqrt{\Lambda Q^2_{\rm m}}}{x^2 Q_{\rm m}} \left[\frac{K}{Q_{\rm m}}-  \phi(x) \right] \,.
\label{E1}
\end{equation}
The electric field changes the sign on the surfaces $x=x_{j}$, for~which
$\phi(x_{j})= \frac{K}{Q_{\rm m}}$.
The typical profiles of the electric field are presented on Figure~\ref{fig5}; these profiles look like the inverted fold-like structure (the fold-like structure will be recovered, if~we change the sign of the magnetic charge $Q_{\rm m}$ and the parameter $K$ simultaneously). On~these profiles one can see two values of $x_j$, in~which the electric field changes the sign. Near~the cosmological horizon, the behavior of this electric field is of the Coulombian type, i.e.,~$E \propto \frac{1}{x^2}$.

\section{Discussion}\label{sec5}

We described an example of a new specific substructure, which can appear in the outer zone of the axionic dyon; we indicated it as a {fold}. The~fold is presented in the profile of the metric function $N(r)$ as a specific zone, which contains the minimum, the~barrier on the left, and~the maximum on the right of this minimum. The~fold is entirely located in the outer zone, i.e.,~it can not be harbored by the event horizon. The~necessary condition of the fold appearance is the inequality $M_1>M>M_{\rm c}$, where $M$ is the asymptotic mass of the dyon, and~$M_1$, $M_{\rm c}$ are the critical masses given by the formulas
\begin{equation}
M_{1} =  \frac{\left(1- \sqrt{1-4 \Lambda Q^2_{\rm m}}+ 4\Lambda Q^2_{\rm m}\right)}{\sqrt{18\Lambda \left(1 - \sqrt{1-4 \Lambda Q^2_{\rm m}} \right)}} \,,
\quad M_{\rm c} = \frac43 |Q_{\rm m}| (\Lambda Q^2_{\rm m})^{\frac14}
\,.
\label{bb2}
\end{equation}
Both of the critical masses $M_1$ and $M_{\rm c}$ contain only two model parameters: the cosmological constant $\Lambda$ and the magnetic charge $Q_{\rm m}$. When $\Lambda=0$ the fold can not be formed, since the maximum on the right of the central minimum disappears. On~the fold bottom the derivative of the gravitational potential vanishes, thus there the massive particle does not feel the gravitational force and can be at rest. The~width and depth of the fold are regulated, according to the Formula (\ref{R32}), by~two dimensionless parameters $\Lambda Q^2_{\rm m}$ and $\frac{M}{M_{\rm c}}$.

Then we analyzed the solution to the key equation for the axion field (\ref{R317}), and~have found that the profile of the axion field reveals the substructure of the same type. To~be more precise,  the~fold-like zones are found in the profile of the function $\phi(x)$, and~in the energy density profile $W(x)$ (see Figures~\ref{fig3} and \ref{fig4}, respectively). In~other words, the~axionic distribution bears the imprint of the fold in the gravitational~potential.

Finally, we studied the profile of the electric field, induced by the magnetic field in the axionic environment. Again, the~fold-like structure has been found in this profile (see Figure~\ref{fig5}). In~more detail, we have seen, that the electric field changes the sign twice in this zone; such a behavior is typical for the double electric layers. Similar results are obtained in the paper~\cite{BG1}, where the change of the electric field direction was associated with the stratification of plasma in the axionic dyon magnetosphere.
We think that development of this idea can be interesting for the procedure of identification of the magnetars based on the fine spectroscopic analysis of obtained data.
Why do we think so? The magnetars possess huge magnetic field of the order $10^{13}$--$10^{15} {\rm G}$; this magnetic field produces spectroscopic effects, such as Zeeman effect~\cite{Zeeman}. If~the dark matter has the axionic nature, the~axionically induced electric field appears in the vicinity of magnetic star. The~corresponding coefficient of transformation is estimated to be less than $10^{-8}$; however, the~axionically induced electric field near magnetars is able to produce the quite distinguishable Stark effect.  When one has both effects in the magnetic and electric fields (parallel and/or crossed), then the possibility appears to combine the  well-elaborated methods and to organize an extended diagnostics~\cite{ZeeStark}.
Clearly, the~standard Coulombian type radial electric field can be produced in many standard charged astrophysical objects, but~if we hope to identify, say, the~axionic dyon, we have to find some very specific detail distinguishing this object. In~this sense the fold-like structure of the profile of the electric field of the axionic dyon, which is described in our work, gives just such a specific detail.
Of course, this idea needs detailed estimations and description of the corresponding diagnostics, however, the~discussion of such questions goes beyond the scope of this~work.

\acknowledgments{The work was supported by Russian Science Foundation (Project No. 16-12-10401), and, partially, by~the Program of Competitive Growth of Kazan Federal University.}

\end{document}